\documentclass[aps,prb,groupedaddress,preprint,showpacs]{revtex4-1}

\usepackage{graphicx}
\usepackage{amsmath}
\usepackage{amssymb}
\usepackage{dcolumn}
\usepackage{hyperref}

\begin{document}

\title{\texorpdfstring{Relaxation of dynamically disordered tetragonal platelets in the relaxor ferroelectric $0.964\mathrm{Na}_{1/2}\mathrm{Bi}_{1/2}\mathrm{TiO}_3-0.036\mathrm{BaTiO}_3$}{Relaxation of dynamically disordered tetragonal platelets in the relaxor ferroelectric NBT-3.6BT}}

\author{Florian Pforr}
\author{Kai-Christian Meyer}
\author{M{\'a}rton Major}\altaffiliation{On leave from Wigner Research Centre for Physics, RMKI, Budapest, Hungary}
\author{Karsten Albe}
\author{Wolfgang Donner}
\email[]{wdonner@tu-darmstadt.de}
\affiliation{Institute of Materials Science, Technische Universit\"at Darmstadt, Alarich-Weiss-Stra\ss e 2, 64287 Darmstadt, Germany}
\author{Uwe Stuhr}
\affiliation{Laboratory for Neutron Scattering and Imaging, Paul Scherrer Institut, 5232 Villigen PSI, Switzerland}
\author{Alexandre Ivanov}
\affiliation{Institut Laue-Langevin, 71 avenue des Martyrs, CS 20156, 38042 Grenoble cedex 9, France}

\date{\today}

\begin{abstract}
The local dynamics of the lead-free relaxor $0.964\mathrm{Na}_{1/2}\mathrm{Bi}_{1/2}\mathrm{TiO}_3-0.036\mathrm{BaTiO}_3$ (NBT-3.6BT) have been investigated by a combination of quasielastic neutron scattering (QENS) and \textit{ab initio} molecular dynamics simulations. In a previous paper, we were able to show that the tetragonal platelets in the microstructure are crucial for understanding the dielectric properties of NBT-3.6BT [F. Pforr \textit{et al.}, Phys. Rev. B 94, 014105 (2016)]. To investigate their dynamics, \textit{ab initio} molecular dynamics simulations were carried out using $\mathrm{Na}_{1/2}\mathrm{Bi}_{1/2}\mathrm{TiO}_3$ with 001 cation order as a simple model system for the tetragonal platelets in NBT-3.6BT. Similarly, 111-ordered $\mathrm{Na}_{1/2}\mathrm{Bi}_{1/2}\mathrm{TiO}_3$ was used as a model for the rhombohedral matrix. The measured single crystal QENS spectra could be reproduced by a linear combination of calculated spectra. We find that the relaxational dynamics of NBT-3.6BT are concentrated in the tetragonal platelets. Chaotic stages, during which the local tilt order changes incessantly on the timescale of several picoseconds, cause the most significant contribution to the quasielastic intensity. They can be regarded as an excited state of tetragonal platelets, whose relaxation back into a quasistable state might explain the frequency dependence of the dielectric properties of NBT-3.6BT in the 100~GHz to THz range. This substantiates the assumption that the relaxor properties of NBT-3.6BT originate from the tetragonal platelets.\end{abstract}

\maketitle

\section{Introduction}
The solid solutions of $\mathrm{Na}_{1/2}\mathrm{Bi}_{1/2}\mathrm{TiO}_3$ (NBT) and BaTiO$_3$, termed NBT-BT for short, have been known for 25 years as lead-free ferroelectric materials.\cite{Takenaka1991} Since lead-free replacements for commonly used lead-based ferroelectrics are a research topic of major interest,\cite{Panda2009, Roedel2009, Shvartsman2012} many attempts have been undertaken to elucidate the structure-property relationships\cite{McQuade2016} and optimize the properties of NBT-BT.\cite{Reichmann2015} Today, NBT-BT-based materials indeed rank among the most promising materials systems for industrial applications\cite{Martinez2013, Kang2015, Taghaddos2015, Song2016} and first products have been commercialized.\cite{Roedel2015}\\

Both NBT and BaTiO$_3$ have $ABO_3$ stoichiometry and crystallize with a perovskite structure with similar lattice parameters. The octahedrally coordinated $B$ sites are occupied by Ti in both cases, whereas the cuboctahedrally coordinated $A$ sites are occupied by Na and Bi in the case of NBT, or Ba in the case of BaTiO$_3$. The solid solution NBT-BT also has a perovskite structure where the $A$ site is shared by Na, Bi, and Ba cations. The addition of a few percent BaTiO$_3$ to NBT has considerable effects on its structure and properties. The crystallographic phase transitions are shifted to significantly lower temperatures.\cite{Hiruma2007, Cordero2010, Zhang2016} A morphotropic phase boundary is found between 6\% and 10\% BaTiO$_3$.\cite{Takenaka1991, Ma2010, Cordero2010, Cordero2014} \textcite{Yao2011} and \textcite{Ma2013} proposed an additional phase boundary around as little as 3\% BaTiO$_3$. Drastic changes of the ferroelectric properties due to the addition of low amounts of barium have also been reported.\cite{Zhang2011, Uddin2014, Kang2015}\\

The temperature dependence of the dielectric permittivity of NBT-BT (below approximately 10\% BaTiO$_3$) is characterized by a broad maximum around 550~K\cite{Greicius2009, Schneider2014} and a pronounced frequency dependence of the permittivity, so that NBT-BT is often described as a relaxor.\cite{Jo2011, Craciun2012, Ge2013, Ahn2016, Lidjici2016} Unlike lead-based relaxors, however, the temperature of the permittivity maximum does not depend on the measurement frequency.\cite{Cowley2011} \textcite{Cross1987} lists three properties of $\mathrm{PbMg}_{1/3}\mathrm{Nb}_{2/3}\mathrm{O}_3$ that are characteristic of relaxor ferroelectrics in general: a strong frequency dispersion of the permittivity, a continuous reduction of ferroelectric behavior upon heating, and a high macroscopic symmetry at low temperature. These properties are commonly attributed to the presence of polar nanoregions,\cite{Tsurumi1994, Toulouse2008, Gehring2012, Hlinka2012} which are described as small regions in which the displacements of the lead cations are aligned.\cite{Bokov2012} This alignment is believed to result from chemical disorder of the $B$-site cations for some relaxors with perovskite structure.\cite{Egami2007, Ohwada2008, Bokov2010} Since the underlying mechanisms have proved to be difficult to probe experimentally, theoretical calculations\cite{Bussmann-Holder2005, Vugmeister2006, Macutkevic2011, Ni2013, Kliem2016} and atomistic simulations\cite{Burton2005, *Burton2006, Grinberg2009, Prosandeev2013, Tomita2013, Takenaka2013, Takenaka2014, Al-Barakaty2015, Prosandeev2016} of lead-based relaxors are widely used. In addition, neutron scattering experiments have been carried out to specifically investigate the relaxation dynamics leading to their frequency-dependent properties.\cite{Hiraka2004, Gvasaliya2004, Gvasaliya2004a, Gvasaliya2005, Gvasaliya2007, Rotaru2008}\\

However, various studies have come to the conclusion that the fundamental mechanisms leading to the relaxor behavior of lead-free materials such as NBT-BT differ from those found in classical lead-based relaxors.\cite{Maurya2013, Schneider2014, Petzelt2015} As a first step towards elucidating the structural origins of the high permittivity of NBT-BT, electric field-dependent features could be identified by means of \textit{in situ} diffuse scattering experiments.\cite{Daniels2011, Daniels2012} The results indicate that the origin of the high permittivity can be found in the atomic structure or on the nanometer scale. Later studies of the microstructure confirmed that the tetragonal platelets\cite{Beanland2011} appear to have a significant influence on the dielectric permittivity of NBT-BT.\cite{Chen2014, Pforr2016, Voegler2017}\\

Additional studies using different experimental methods have provided further clues concerning the intimate connection between the octahedral tilt symmetries and the ferroelectric properties in NBT-based materials.\cite{Ma2010, Jo2011, Rao2013} This can be understood given that the local symmetry, which is mainly determined by the tilt system, directly limits the possible polarization directions.\cite{Jaffe1971, Ogino2014} It has also been shown that the first coordination shell of the bismuth cations, which carry the largest part of the polarization, is tilt-dependent,\cite{Rao2016} and Bi-O vibrations make a major contribution to the dielectric response.\cite{Niranjan2013} Finally, several studies have provided indications of coupling between phonon modes related to octahedral tilting or ferroelectricity,\cite{Groeting2013, Mihailova2013, Cordero2014} even if this coupling may not be very strong.\cite{Benedek2013, Kitanaka2016}\\

As indicated above, another fundamental aspect of the structure of perovskite relaxors is the question whether and how cations sharing the same lattice site are ordered. The direct influence of the cation order on the dielectric properties has been clearly demonstrated for lead-based relaxors.\cite{Cross1987, Tinte2006, Bokov2010} In most cases, only short-range order exists.\cite{Cross1987, Cross1995, Zhang1996, Hlinka2006, Goossens2013} The most popular model features small islands of ordered cations, embedded in a matrix with complete cation disorder. Such short-range order has been found to influence $A$-site displacements\cite{Hiraka2004, Tinte2006, Maier2011, Prosandeev2016} and thus also phase transition temperatures.\cite{Prosandeev2016}\\

In the case of NBT, the coexistence of cation order and disorder has also been observed,\cite{Petzelt2004, Isupov2005} although other studies have concluded that the cations possess long-range order\cite{Pan2016} or no order at all.\cite{Aksel2013, Dawson2015} In analogy with lead-based relaxors, the model of randomly arranged, chemically ordered regions in a disordered matrix is often used.\cite{Groeting2011, Groeting2012} Similar models have also been found for the closely related compounds $\mathrm{K}_{1/2}\mathrm{Bi}_{1/2}\mathrm{TiO}_3$\cite{Jiang2017} and NBT-BT.\cite{Yunfei2012} Just like in NBT, the discussion about the cation order of NBT-BT is still open, since \textcite{Kling2010} could not identify any cation order experimentally. Different properties of NBT-BT have been explained with reference to the $A$-cation disorder (as opposed to long range order), most importantly the dielectric dispersion\cite{Gomah-Pettry2004, Xu2008a} and the diffuse phase transition.\cite{Lidjici2016}\\

\begin{figure*}
\includegraphics[width=17.0cm]{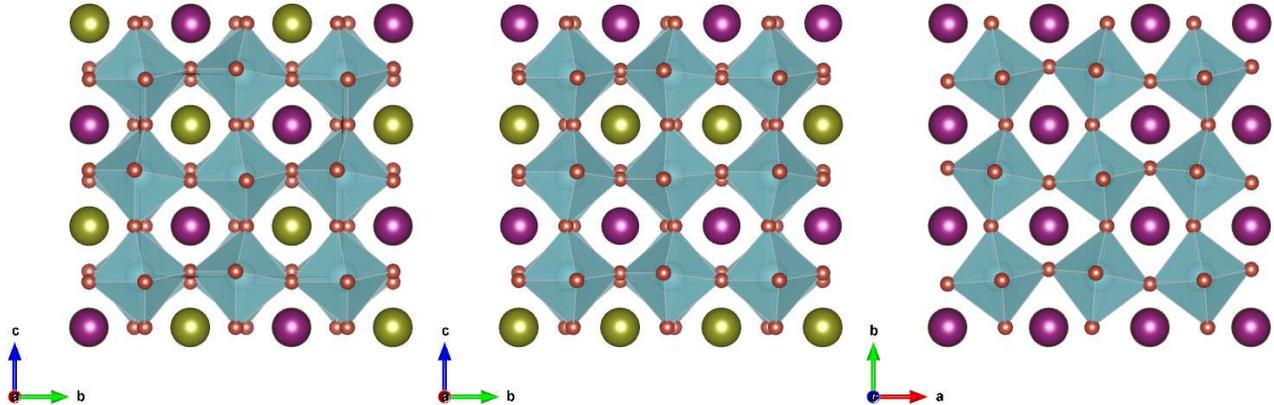}%
\caption{Ground state structures of NBT with 111 cation order viewed along the $a$ axis (left) and with 001 cation order viewed along the $a$ and $c$ axes (middle and right, respectively). The structures were drawn using VESTA\cite{Momma2011} with the following color scheme: Na$^+$: yellow, Bi$^{3+}$: violet, Ti$^{4+}$: light blue, O$^{2-}$: red.\label{fig:DFT-structures}}
\end{figure*}

Since the octahedral tilting is central to this study, a possible influence of the $A$-cation order also needs to be considered. It has already been demonstrated by first-principles simulations that depending on the cation order, different tilt systems are energetically favorable.\cite{Groeting2011, Groeting2012} Specifically, the ground state tilt structure of 111-ordered NBT was found to be a$^-$a$^-$a$^-$ (Glazer notation\cite{Glazer1972, *Glazer1975}), whereas the ground state of 001-ordered NBT is a$^-$a$^-$c$^+$. These two structures are shown in Fig.\nolinebreak\ \ref{fig:DFT-structures}. Here, we denote the rock salt order of the $A$-cations as 111-order and the alternating stacking of Na- and Bi-layers along the $c$ axis as 001-order. It should be noted that superlattice reflections in scattering experiments may not only result from octahedral tilting, but also from long-range cation order.\cite{Levin2012} However, such cation-order induced superlattice reflections have not been observed in NBT-BT. Bearing the cation-order dependence of the ground state tilt structure in mind, it appears reasonable to assume that the coexistence of tetragonal platelets and a rhombohedral matrix in NBT-BT results from locally different cation order, as proposed in Ref.\nolinebreak\ \onlinecite{Pforr2016}. Further simulation studies have confirmed that the cation order has a significant impact on the tilt defect energies and tilt dynamics, too.\cite{Meyer2015, Meyer2017}\\

The dynamics of NBT have previously been characterized using quasielastic neutron scattering (QENS)\cite{Vakhrushev1982, Vakhrushev1983, Vakhrushev1985} and inelastic light scattering,\cite{Siny2000, *Siny2001, Jackson2014, Fedoseev2015} mostly to elucidate its phase transition mechanisms. Broadband dielectric spectroscopy has been used to identify some relaxation modes in NBT,\cite{Petzelt2004, Petzelt2014, Petzelt2015} but a clear-cut atomistic explanation for its relaxor properties has not yet been presented.\\

This study combines QENS and \textit{ab initio} molecular dynamics (MD) simulations to shed more light on the atomistic origins of the relaxor properties of $0.964\mathrm{Na}_{1/2}\mathrm{Bi}_{1/2}\mathrm{TiO}_3-0.036\mathrm{BaTiO}_3$ (NBT-3.6BT). In most applications, QENS measurements are used to probe hydrogen dynamics. Since the predominant isotope, $^1$H, is a predominantly incoherent scatterer, the observed incoherent QENS results from dynamic autocorrelations of the hydrogen atoms. On the other hand, coherent QENS can be observed in samples containing predominantly coherent scatterers.\cite{Hempelmann2000} It results from both dynamic pair-correlations and dynamic autocorrelations. Of course, coherent and incoherent QENS can also be superimposed if a mixed scatterer exhibits dynamic correlations or if both coherent and incoherent scatterers exhibit dynamic correlations. In NBT-BT, bismuth, barium, and oxygen are predominantly coherent scatterers, whereas sodium and titanium are mixed scatterers. This means that in principle, coherent and incoherent QENS could be observed, depending on the relevant dynamic correlations.\\

For investigating the dynamic properties of NBT and NBT-BT, classical MD simulations cannot be used, since no interatomic potentials exist. Instead, we use \textit{ab initio} MD simulations.\cite{Marx2009} In this method, all interatomic forces are computed from first-principles on a frame-by-frame basis.\cite{Car1985, Payne1992, Kresse1993} Consequently, specific features such as mixed ionic and covalent bonds, and quantum mechanical effects are implicitly accounted for. The downside is the high computational cost, which is a direct consequence of the computation technique. Whereas in classical MD simulations atoms are the only entities, in \textit{ab initio} MD simulations the valence electrons determine the forces on the atoms, thus increasing the number of interacting particles in the simulation. Since the number of valence electrons is particularly high for titanium and bismuth, \textit{ab initio} MD simulations of NBT have a high computational demand. Conversely, the investigation of dynamic properties requires a high number of simulation steps, which can only be achieved by reducing the size of the simulation box. A box of $2\times 2\times 2$ single perovskite unit cells ($a\approx 3.9~\mathrm{\mathring{A}}$) with periodic boundary conditions is large enough to adopt different octahedral tilt orders. Still, this size is far too small for simulations featuring complex cation orders, barium doping, coexistence of multiple tilt systems, or ensembles of polar nanoregions. Despite these limitations, suitable model calculations have revealed cation-order specific and temperature-dependent dynamic behavior, which was subsequently interpreted with respect to the more complex microstructure of NBT-BT.\\

\section{Experimental and computational methods}
QENS spectra between 310~K and 780~K were collected at EIGER\cite{Stuhr2017} (PSI, Switzerland) using a small furnace as described in Ref.~\onlinecite{Pforr2016}. Spectra from 10~K to 300~K were collected at IN8\cite{Hiess2006} (ILL, France) using the same $0.964\mathrm{Na}_{1/2}\mathrm{Bi}_{1/2}\mathrm{TiO}_3-0.036\mathrm{BaTiO}_3$ single crystal sample,\cite{Pforr2016} mounted in an orange cryostat using bent Al sheet and VGE-7031 varnish (Cryophysics GmbH, Darmstadt). The incident and scattered wavevectors were defined using a Si monochromator and double focusing pyrolytic graphite analyzer.\\

QENS spectra were measured at one superlattice reflection that is due to in-phase tilting [$\mathbf{Q} = \frac{1}{2}(310)$] and one that is due to anti-phase tilting [$\frac{1}{2}(311)$].\cite{Pforr2016} Reciprocal space coordinates are given with respect to the single perovskite unit cell with $a\approx 3.9~\mathrm{\mathring{A}}$. Nine data sets above and five at or below room temperature (RT) were collected at each $\mathbf{Q}$ point. Additional background scans were carried out at 310~K and 780~K, using only the furnace and empty container. The background correction of the data sets above RT was performed using linearly interpolated background intensities. If the background contribution to any given data point was greater than half the intensity with sample, this data point was excluded from further treatment.\\

\textit{Ab initio} MD calculations were performed on NBT within the density functional theory Vienna Ab initio Simulation Package (VASP).\cite{Kresse1996, *Kresse1996a} Projector augmented waves\cite{Kresse1999, Bloechl1994} together with the local density approximation were applied.\cite{Perdew1981} The cell consisted of 40 atoms and the valence electrons configurations were: O:~$2s^2 2p^4$, Na:~$2p^6 3s^1$, Ti:~$3s^2 3p^6 4s^23d^2$, Bi:~$5d^{10} 6s^2 6p^3$. The plane wave energy cut-off was set to 500~eV. Electronic configurations were optimized until the energy of the cell was changing by less than $10^{-5}$~eV. The number of k-points was set to one. A time step of 1~fs was used and time spans of 25~ps to 100~ps were simulated in a canonical ensemble. Two different $A$-cation orders were studied: 001 and 111 order. The initial structure was chosen to be the rhombohedral phase with an $a^-a^-a^-$ tilt pattern (space group $R3c$). The space group is given for the high symmetry configuration (no $A$-cation order).\\

The evaluation of the \textit{ab initio} MD data was performed using different versions of the nMoldyn\cite{Hinsen2012} suite. The root mean square displacement (RMSD) quantifies the difference between the initial structure and the snapshot structure at any given time step. The static coherent structure factor (SCSF) is the component of the scattering function $\mathbf{S}(q,\omega)$ that results from elastic coherent neutron scattering. It is the neutron scattering counterpart of the x-ray structure factor. Both the RMSD and the SCSF were calculated using nMoldyn 3.0.8. The dynamic coherent structure factor (DCSF) includes both elastic and inelastic coherent neutron scattering. It was calculated using nMoldyn 4.0.0 using an energy resolution of 0.0688~ps$^{-1}$ (standard deviation), which corresponds to the instrumental resolution of EIGER (full width at half maximum: 0.67~meV). The dynamic incoherent structure factor includes the elastic and inelastic incoherent neutron scattering. It was also calculated, but turned out to be negligible compared to the DCSF.\\

\section{Results and Discussion}
\subsection{Analysis of QENS data}
\begin{figure}
\includegraphics[width=8.5cm]{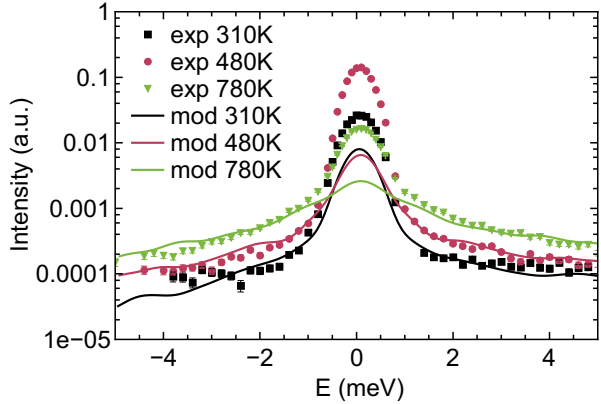}%
\caption{Comparison of experimental and modeled QENS spectra at $\frac{1}{2}(310)$ above room temperature. The quasielastic component is reproduced, but not the elastic component. Each model spectrum was synthesized by linear combination of four data sets with different cation orders and temperatures. See the text for details. To compensate for the effects of barium doping in the measured single crystal, the temperature scale was empirically shifted according to Eq.\protect\nolinebreak\ (\ref{eq:exptomod}).\label{fig:expt_vs_model_highT}}
\end{figure}

In Fig.\nolinebreak\ \ref{fig:expt_vs_model_highT}, experimental and modeled QENS spectra at the reciprocal lattice point $\frac{1}{2}(310)$ are compared. DCSF calculations were carried out on \textit{ab initio} MD simulations with pure 001 or 111 order in the simulation temperature range $100~\mathrm{K}\leq T_\mathrm{sim}\leq 1000~\mathrm{K}$. Linear combinations of the 001 and 111 spectra at a given $T_\mathrm{sim}$ were calculated to represent a macroscopic sample. The resulting spectra were again combined linearly to give one calculated spectrum at the model temperature $T_\mathrm{mod}$ with $T_{\mathrm{sim}, 1}\leq T_\mathrm{mod}\leq T_{\mathrm{sim}, 2}$. For example, the spectrum at $T_\mathrm{mod} = 770~\mathrm{K}$ (corresponding to $T_\mathrm{exp} = 480~\mathrm{K}$, see below) is the sum of 30\% of the $T_{\mathrm{sim}, 1} = 700~\mathrm{K}$ spectrum and 70\% of the $T_{\mathrm{sim}, 2} = 800~\mathrm{K}$ spectrum.\\

The QENS spectra were measured on NBT-3.6BT at $\frac{1}{2}(310)$ and $\frac{1}{2}(311)$, with the experiment temperature $T_\mathrm{exp}$ in the range $310~\mathrm{K}\leq T_\mathrm{exp}\leq 780~\mathrm{K}$. The quasielastic part of the spectra was compared with respect to their profiles and relative intensities. The closest match was achieved using a 001-ordered fraction of 90\% and a 111-ordered fraction of 10\%. Furthermore, the temperature range had to be rescaled according to
\begin{equation}
T_\mathrm{mod} = 0.5574\cdot T_\mathrm{exp} + 498.9~\mathrm{K}.\label{eq:exptomod}
\end{equation}

With a common intensity scaling factor as the only other free parameter, the quasielastic part of 9 spectra each at $\frac{1}{2}(310)$ and $\frac{1}{2}(311)$ was accurately reproduced. The main reason for the shift of the temperature scale is the influence of the barium addition on the transition temperatures of NBT. The uncertainty of the MD temperatures and effects of the limited box size may also play a role. Further possible reasons for discrepancies between experimental and \textit{ab initio} results are discussed in Ref.\nolinebreak\ \onlinecite{Groeting2014}. In contrast to the quasielastic part of the spectra, the elastic part is not reproduced satisfactorily. This is due to the highly complex microstructure of real NBT-BT, which cannot be accounted for in the simplified model. For instance, crystallographic defects lead to diffuse scattering, which is also visible as part of the elastic signal in the QENS spectra. In the case of NBT-3.6BT, the diffuse streaks running through $\mathbf{Q} = \frac{1}{2}(310)$ and $\frac{1}{2}(311)$\cite{Pforr2016} probably account for most of the difference between the modelled and experimental spectra.\\

In terms of octahedral tilting, we regard the 001 ordered NBT as a model system for the tetragonal platelets in NBT-3.6BT, since both are characterized by preferred c$^+$ tilting. By the same reasoning, we regard the 111 ordered NBT as a model system for the rhombohedral matrix due to the preferred a$^-$a$^-$a$^-$ tilt order. Consequently, the fits shown in Fig.\nolinebreak\ \ref{fig:expt_vs_model_highT} appear to imply a volume fraction of 90\% tetragonal platelets, which would contradict the results published previously in Ref.\nolinebreak\ \onlinecite{Pforr2016}. This apparent contradiction can only be resolved by the assumption that the rhombohedral matrix is less dynamic and contributes much less to the QENS of NBT-3.6BT than it would in an ideal 111-ordered NBT crystal. Since the real crystal contains a mixture of multiple phases on the length scale of a few nanometers,\cite{Pforr2016} none of the single-phase domains possess the same long-range periodicity as the model structures. In addition, the rhombohedral-tetragonal interfaces may be much less dynamic than the bulk structures and thus lead to a clamping effect in the real crystal. Both effects suppress phonon dynamics and thus relatively enhance incoherent dynamics, particularly in the smaller tetragonal domains. This justifies the higher tetragonal phase fraction in the simplified model.\\

\begin{figure}
\includegraphics[width=8.5cm]{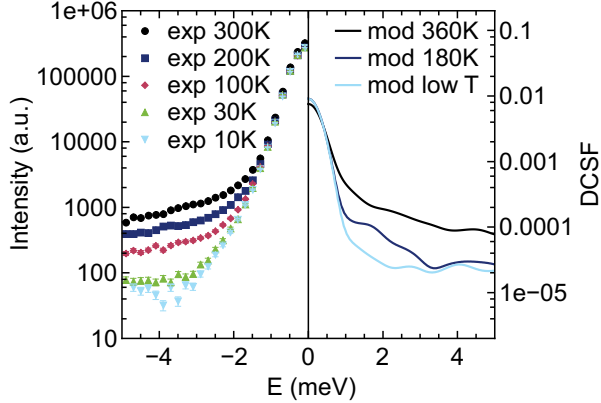}%
\caption{Comparison of experimental\cite{Pforr2013} (left) and modeled (right) QENS spectra at $\frac{1}{2}(310)$ below room temperature. The $T_\mathrm{exp}$ values given on the right hand side were extrapolated using Eq.\protect\nolinebreak\ (\ref{eq:exptomod}). The label `low T' is used in place of the calculated value of 0~K due to the uncertainty associated with the extrapolation. The pronounced oscillations of the calculated DCSF are mainly due to the finite simulation time and size of the simulation box.\label{fig:expt_vs_model_lowT}}
\end{figure}

Fig.\nolinebreak\ \ref{fig:expt_vs_model_lowT} extends the comparison range from Fig.\nolinebreak\ \ref{fig:expt_vs_model_highT} to temperatures below RT. Since the experimental data could not be corrected for background, and the resolution functions do not match, a quantitative one-to-one comparison is not possible in this case. Nevertheless, the quasielastic part of the spectra appears to match qualitatively. This indicates that the mathematical model containing 90\% 001-ordered NBT also exhibits the same dynamics as NBT-3.6BT below RT, while confirming the previously discussed temperature shift between $T_\mathrm{exp}$ and $T_\mathrm{mod}$. The measured QENS intensity change between 300~K and 10~K of about one order of magnitude and the change of the modeled intensity between 360~K and low temperature match rather well. Furthermore, Eq.\ (\ref{eq:exptomod}) appears to be applicable not only above RT, i.e., in the temperature range for which it was originally devised, but also below RT, thus confirming the validity of this approach.\\

\begin{figure*}
\includegraphics[width=13.0cm]{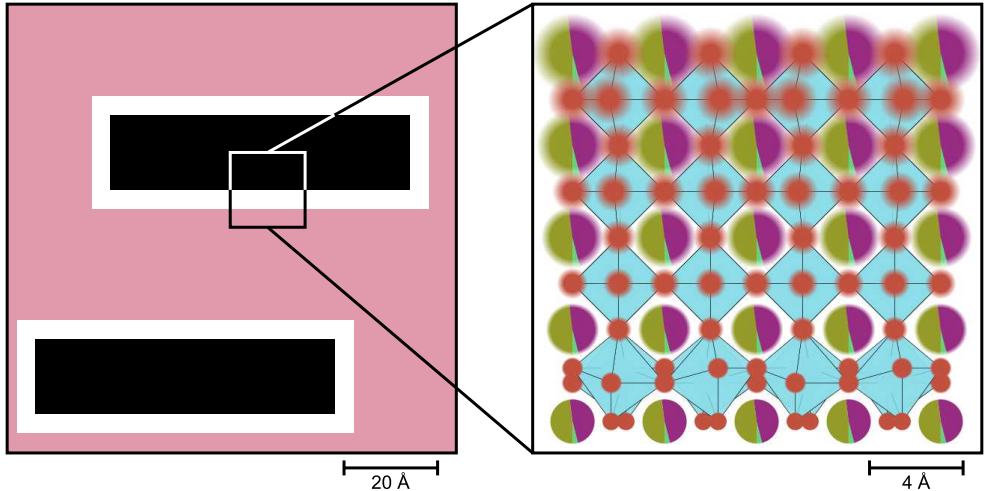}%
\caption{Location of the dynamic effects in the nanostructure of NBT-3.6BT. The tetragonal regions (black) appear to be much more dynamic than the rhombohedral ones (light red), so that we suspect the most pronounced dynamics in the center of the tetragonal platelets. The white regions represent the cubic intermediate phase as in Ref.\protect\nolinebreak\ \onlinecite{Pforr2016}. The color scheme for the atoms in the enlarged part corresponds to that in Fig.\protect\nolinebreak\ \ref{fig:DFT-structures}. The added Ba$^{2+}$ is shown in light green. Since the actual cation arrangement is not known, a random distribution is shown instead.\label{fig:Platelet}}
\end{figure*}

As discussed above, the most significant contribution to the measured NBT-3.6BT QENS spectra stems from the tetragonal platelets. Since their dynamics can be more easily excited thermally, we can also assume that outside influences such as an external electric field lead to a particularly strong response. The concentration of the local dynamics is schematically depicted in Fig.\nolinebreak\ \ref{fig:Platelet}. We assume that the rhombohedral matrix is not only comparatively static, but also reduces the dynamics of the tetragonal platelets close to the interface. The dynamics of the tetragonal platelets would thus increase with increasing distance from the interface. Recalling the comparable depth dependence of the dielectric permittivity devised in Ref.\nolinebreak\ \onlinecite{Pforr2016}, a possible correlation between the dynamics of the tetragonal platelets and the dielectric permittivity can be postulated.\\

\subsection{Interpretation of \textit{ab initio} MD data}

\begin{figure*}
\includegraphics[width=17.0cm]{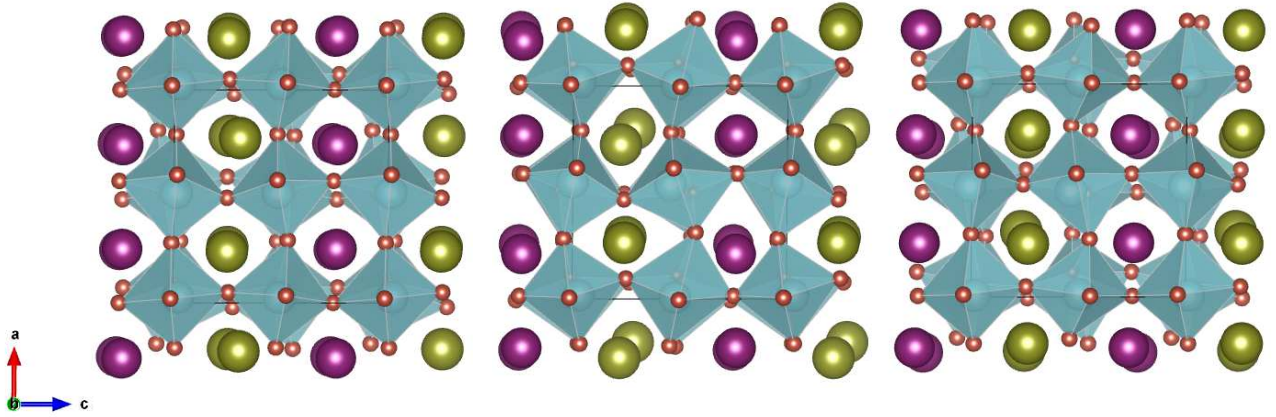}%
\caption{Snapshot structures of 001-ordered NBT at the simulation temperature $T_\mathrm{sim}=700~\mathrm{K}$, viewed along the $b$ axis. Snapshots were taken at 40900~fs (left), 41130~fs (middle), and 41320~fs (right), and drawn using VESTA\cite{Momma2011} (same color scheme as in Fig.\protect\nolinebreak\ \ref{fig:DFT-structures}). Rapid changes of the local $b$-axis tilting from anti-phase via in-phase back to anti-phase are clearly seen. Concurrently, the tilt angle about the $c$ axis increases between 40900~fs and 41130~fs. Subsequently, the $a$-axis tilting changes from nearly tilt-free at 41130~fs to strongly anti-phase at 41320~fs.\label{fig:Tilt-chaos}}
\end{figure*}

In the following, the structural evolution of the \textit{ab initio} MD data over the course of the simulations will be analyzed in more detail. At first glance, changes on the time scale of 100~fs become apparent. These include thermal oscillations and single tilt flips. At second glance, patterns in these quick changes become apparent, which typically change on the time scale of a few picoseconds. Phases of the simulation with one such pattern are hereafter referred to as `stages'. The data set with $T_\mathrm{sim} = 700~\mathrm{K}$ and 001 order is chosen as a representative example for this analysis. Fig.\nolinebreak\ \ref{fig:Tilt-chaos} shows snapshots from the \textit{ab initio} MD calculation at three different time steps. Whereas the first and third snapshots show anti-phase octahedral tilting about the $b$ axis, the second snapshot shows in-phase tilting. Additional changes of the tilting about the $a$ and $c$ axes are also seen in this projection. All of these changes occur within a few hundred femtoseconds, and many more precede and succeed them at a similar rate. However, these chaotic changes of the tilt order must not be confused with the more regular thermal oscillations of stable average structures, which occur during other stages of the simulation. The latter, termed quasistable stages, are sometimes separated by single tilt flips. Regarding the free energy landscape, chaotic stages indicate that the activation barrier between the observed tilt structures is smaller than the thermal energy. This means that the system fluctuates between numerous metastable configurations during chaotic stages, because the local energy landscape around the occurring configurations is rather flat. Once the system finds a deeper energy minimum, the thermal energy only leads to thermal oscillations of this new quasistable structure.\\

\begin{figure}
\includegraphics[width=8.5cm]{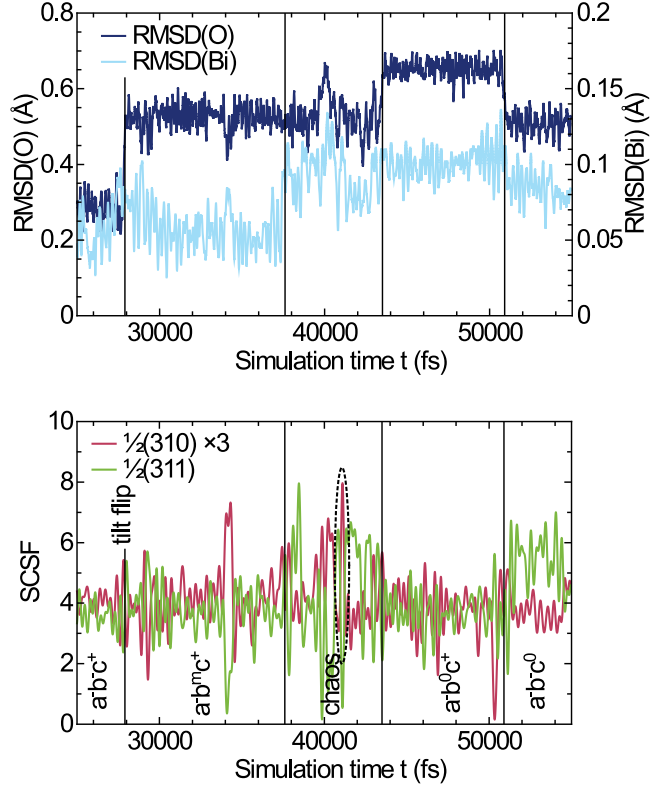}%
\caption{Development of the octahedral tilt structure in the \textit{ab initio} MD simulation with 001 cation order at $T_\mathrm{sim} = 700~\mathrm{K}$ (corresponding to $T_\mathrm{exp} = 360~\mathrm{K}$). Top: RMSD of the oxygen and bismuth atoms. Vertical lines indicate stages separated by sudden jumps of the RMSD. Bottom: SCSF indicate the degree of in-phase [$\frac{1}{2}(310)$; scaled up by a factor of 3] and anti-phase [$\frac{1}{2}(311)$] tilt order. SCSF data were obtained on a frame-by-frame basis and subsequently smoothed using an FFT filter to reduce high frequency oscillations. Also indicated are tilt systems in stages with stable tilt symmetry, as determined from one snapshot per stage. `m' indicates mixed tilting. The dotted ellipse highlights the period during which the snapshots in Fig.\protect\nolinebreak\ \ref{fig:Tilt-chaos} were taken.\label{fig:RMSD+SCSF}}
\end{figure}

The upper graph in Fig.\nolinebreak\ \ref{fig:RMSD+SCSF} depicts the development of the RMSDs of oxygen and bismuth. Different stages are separated by black, vertical lines. At the transition from one stage to the next, the oxygen sublattice often rearranges cooperatively. The middle stage around 40000~fs is characterized by enhanced fluctuations of the oxygen order. In many cases, correlations between oxygen rearrangements and bismuth rearrangements can be seen. This appears to confirm the strong interaction between the dielectric polarization, mainly carried by the bismuth cations, and the tilting of the oxygen octahedra.\\

The lower graph in Fig.\nolinebreak\ \ref{fig:RMSD+SCSF} depicts the corresponding evolution of the $\frac{1}{2}(310)$ (magnified by 3) and $\frac{1}{2}(311)$ SCSF. They are effectively order parameters for in-phase and anti-phase tilting, respectively. In addition, the octahedral tilt structure of each stage is specified, as observed in one snapshot per stage. It is clearly seen that jumps in the RMSD often correspond to instantaneous changes of the octahedral tilt symmetry. Stages with constant RMSD exhibit constant tilt symmetry. Stages with pronounced RMSD fluctuations are chaotic in the sense that the tilt symmetry changes almost incessantly. The transition from the first to the second stage, both of which are quasistable, corresponds to a single tilt flip, leading to a shift of the oxygen atoms, but no significant change in symmetry. Apart from this tilt flip and a brief spike around 34000~fs, only thermal oscillations about the stable structure occur. Like in the RMSD, the middle stage exhibits much more frequent symmetry changes. A quasistable average structure cannot be observed. Due to this dynamic nature, this stage is termed `chaotic'. The snapshots in Fig.\nolinebreak\ \ref{fig:Tilt-chaos} were taken in the highlighted region. At the end of this stage, the system finds a new stable structure with the same symmetry, but different arrangement of the atoms compared to previous stages. This is clearly demonstrated by the RMSD. The last rearrangement of the oxygen atoms shown in this figure correlates with a steplike increase in antiphase tilt order.\\

Around 38000~fs, the system is thermally excited from a quasistable state into a chaotic state. Despite developing a number of different structures, the system only relaxes into another quasistable structure after a significant length of time. The duration of the chaotic stage can thus be regarded as a relaxation time. It gives an indication of how much time the system may need to relax into a stable configuration after a change of the free energy landscape due to an outside influence, e.g., a change in electric field. Relaxation times on the order of 1~ps to 10~ps would lead to a frequency dependence of the dielectric response above 100~GHz, as observed for example in Ref.\nolinebreak\ \onlinecite{Petzelt2004, Petzelt2014}.\\

Above $T_\mathrm{exp}\approx 360~\mathrm{K}$, faster switching between different stages is observed along with a higher time fraction of chaotic stages. At $T_\mathrm{exp}\approx 720~\mathrm{K}$, the thermal energy is higher than the stabilization of any ordered state, so that no quasistable stages occur. Instead, the system remains in the chaotic state.\\

\begin{figure*}
\includegraphics[width=15.0cm]{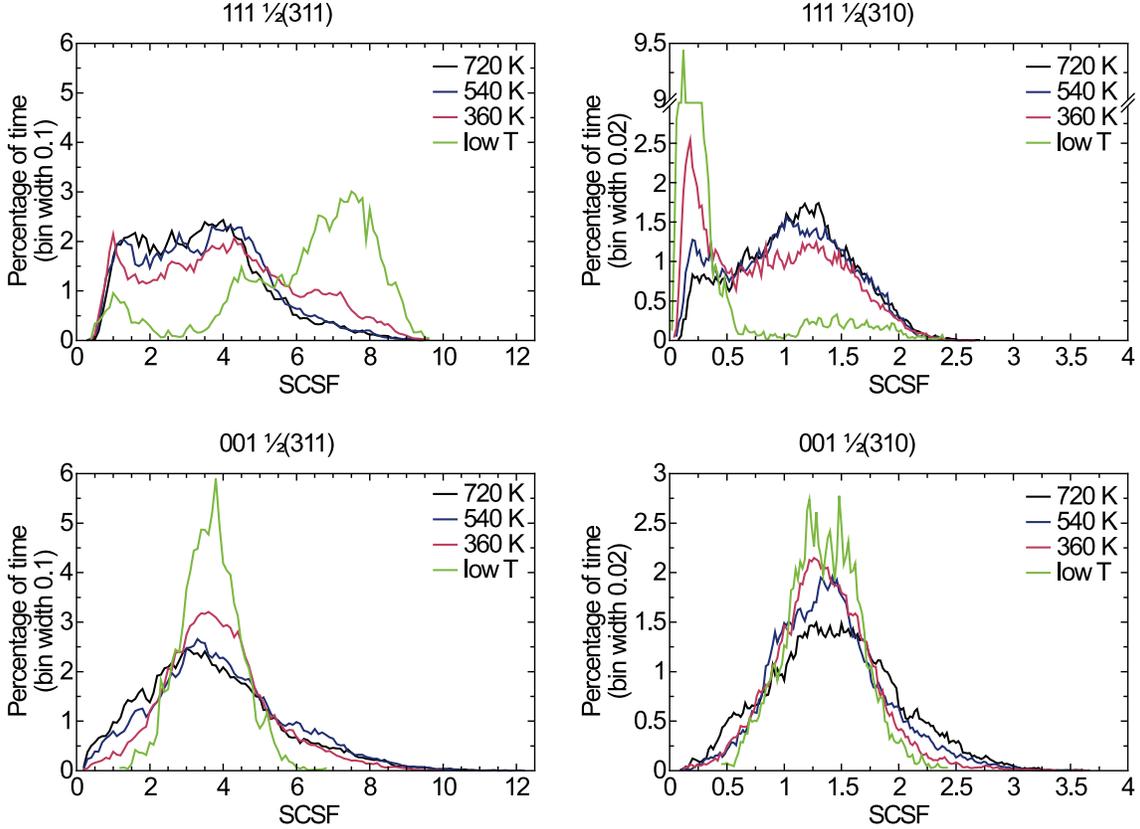}%
\caption{Time-dependence of the SCSF (abscissa) condensed into histograms. Left: $\frac{1}{2}(311)$, right: $\frac{1}{2}(310)$; top: 111 order, bottom: 001 order. Temperatures are given as $T_\mathrm{exp}$, calculated using Eq.\protect\nolinebreak\ (\ref{eq:exptomod}).\label{fig:SCSF_Hist_R+T}}
\end{figure*}

In order to gain insights into the temperature and cation order dependence of the SCSF, histograms of the time-dependent (frame-by-frame) SCSF were calculated. They are summarized in Fig.\nolinebreak\ \ref{fig:SCSF_Hist_R+T}. The amount of anti-phase tilt order is represented by the $\frac{1}{2}(311)$ SCSF on the left, whereas the $\frac{1}{2}(310)$ SCSF on the right represents the amount of in-phase tilt order. The histograms in the top and bottom rows were derived from simulations with 111 and 001 cation order, respectively.\\

Switching between different quasistable states at low temperature leads to multimodal distributions for the SCSF of 111-ordered NBT. At higher temperatures, the distributions become very broad and the differences between 001 and 111 order are continuously reduced. The fact that c$^+$ tilting is favored by 001 order makes the influence of the cation order apparent. It leads to a shift of the mean $\frac{1}{2}(310)$ SCSF to higher values. On the other hand, 111 cation order leads to a more frequent disappearance of either tilt order. These histograms thus confirm the suppression of in-phase tilting by 111 cation order, whereas states without any in-phase tilt component hardly occur in 001-ordered NBT. Similarly, a higher degree of anti-phase tilting is observed in 111-ordered NBT than in 001-ordered NBT. This can be seen as confirmation that 111-ordered NBT behaves more like the rhombohedral matrix in NBT-3.6BT, whereas the tilt structure of the 001-ordered NBT matches that of the tetragonal regions more closely.\\

With increasing temperature, the distribution of both SCSF broadens in the case of 001-ordered NBT. On the other hand, a pronounced shift is observed in the case of 111-ordered NBT, in which anti-phase order is reduced while in-phase order increases. Furthermore, the change in the nature of the dynamics in 111-ordered NBT is clearly seen. Below RT, separate, well-defined states exist, in which the system remains for longer periods of time during the quasistable stages. Transitions between the quasistable states are so quick that the total duration of the intermediate states is significantly shorter. At higher temperatures, chaotic stages occur and account for increasing fractions of the simulation time, leading to broader SCSF distributions. The increasing similarity of the distributions for 001- and 111-ordered NBT clearly shows that the influence of the cation order becomes less significant, although the preference of c$^+$ tilting in 001-ordered NBT is still visible at 720~K. In the high temperature limit, both systems tend to form a highly dynamic, isotropic phase, which can be identified with the paraelectric (average cubic) phase of NBT-3.6BT. This isotropic phase should not be confused with the much discussed isotropic point that occurs in the antiferroelectric region, i.e., at much lower temperatures.\cite{Isupov2005}\\

\begin{figure}
\includegraphics[width=8.5cm]{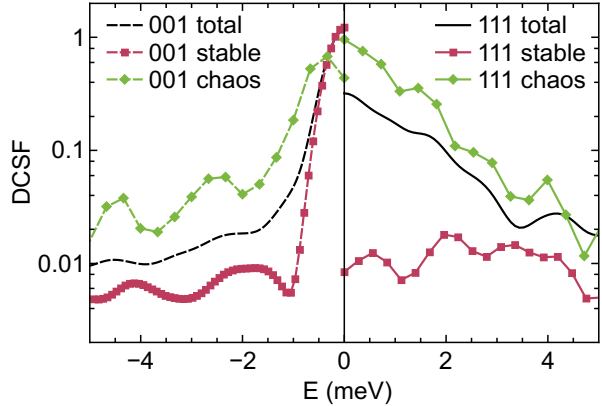}%
\caption{DCSF at $T_\mathrm{sim} = 700~\mathrm{K}$ ($T_\mathrm{exp} = 360~\mathrm{K}$) with 001 cation order (left) and at $T_\mathrm{sim} = 600~\mathrm{K}$ ($T_\mathrm{exp} = 180~\mathrm{K}$) with 111 order (right), at $\frac{1}{2}(310)$. The total DCSF, which was used in the reproduction of the measured QENS (compare Fig.\protect\nolinebreak\ \ref{fig:expt_vs_model_highT}), is shown as a black line. Additionally, the individual contributions from chaotic and quasistable stages, as identified in Fig.\protect\nolinebreak\ \ref{fig:RMSD+SCSF}, are shown as lines with red squares and green diamonds, respectively (dark and light grey in the printed version).\label{fig:DCSF_310_111vs001}}
\end{figure}

The different components of the DCSF are shown in Fig.\nolinebreak\ \ref{fig:DCSF_310_111vs001}. The most precise separation of the quasistatic and chaotic stages could be achieved for the $T_\mathrm{sim} = 700~\mathrm{K}$ dataset (corresponding to $T_\mathrm{exp} = 360~\mathrm{K}$) of 001-ordered NBT, shown on the left, and for the $T_\mathrm{sim} = 600~\mathrm{K}$ dataset ($T_\mathrm{exp} = 180~\mathrm{K}$) dataset of 111-ordered NBT, shown on the right. It is clearly seen that only the chaotic stages contribute significantly to the quasielastic scattering on the relevant timescale of about 1~ps to 4~ps. The scattering from the quasistatic stages is purely elastic. The total DCSF are the time-weighted averages of the respective chaotic and quasistatic contributions. This means that on the timescale probed by our QENS experiment, relaxational motions only occur during the chaotic stages. Since such relaxations must be regarded as the origin of the frequency-dependent behavior of NBT-3.6BT in the range of 100~GHz, these chaotic relaxations provide the most natural explanation for the relaxor properties.\\

\begin{figure}
\includegraphics[width=8.5cm]{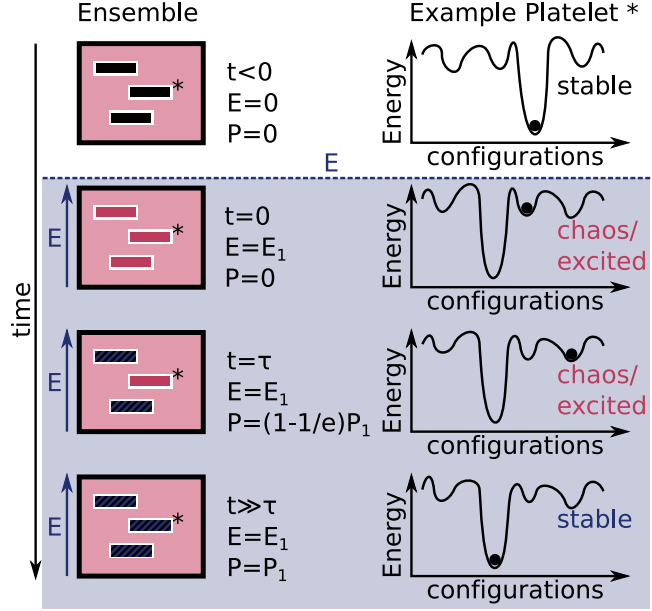}%
\caption{Proposed fast switching behavior of an ensemble of tetragonal platelets (left) and an example platelet (right), marked by an asterisk, upon application of an electric field at $t=0$. Initially, most platelets are in a stable structural configuration, as indicated by the black color. Light red and white regions represent rhombohedral and cubic regions as in Fig.\protect\nolinebreak\ \ref{fig:Platelet}. The application of an electric field changes their free energy landscapes, so that their configurations become metastable. Subsequently, the thermal energy is sufficient for inducing chaotic configuration changes in the excited state of the platelets (red). With time, more and more platelets encounter a configuration that is stabilized by the applied electric field (hatched black and blue). They carry a nonzero polarization that is consistent with the external field. The ensemble average time for one platelet to reach a stable configuration through a chaotic stage thus defines the relaxation time of the macroscopic polarization.\label{fig:Platelet-relaxation}}
\end{figure}

Now let us consider the application of an electric field (several kilovolts per millimeter) to the system. We believe that the field will change the local energy landscape, particularly in the tetragonal platelets, and thus destabilize the previously quasistable structure. This may induce a chaotic stage until a new quasistable state is encountered. Since the new quasistable structure is induced by the applied electric field, it probably carries a nonzero polarization that is aligned with the applied field. This mechanism is sketched in Fig.\nolinebreak\ \ref{fig:Platelet-relaxation}. At high temperatures, however, no quasistable states exist. This means that even though the system will react to the applied field and carry an average polarization during the chaotic stage, it will remain dynamic and this polarization will collapse as soon as the field is removed. This is characteristic of a paraelectric response.\\

Our hypothesis of the dynamic response of NBT-3.6BT to an applied electric field is depicted in Fig.\nolinebreak\ \ref{fig:Platelet-relaxation}. On the left, the reaction of an ensemble of tetragonal platelets to the application of an electric field is shown schematically. On the right, the current configuration of an example platelet is shown within the (dynamic) free energy landscape. On the left, the example platelet is marked by an asterisk. At $t=0$, an electric field is applied. This changes the free energy landscape of the tetragonal platelets, so that they enter a chaotic stage. This can also be thought of as an excited state. Successively, the platelets adopt a new configuration, which has become quasistable due to the applied field. Finally, the entire sample carries a polarization corresponding to the external electric field.\\

The time it takes for any given tetragonal platelet to encounter a new quasistable state varies, partly due to the local cation arrangement and size of the platelets. As a result, the macroscopic response of the sample builds up continuously until it saturates. The time dependence of this saturation behavior reflects the distribution of characteristic times of all platelets. The most straightforward model to describe such saturation behavior is the relaxation following an exponential law: $P\propto 1-e^{-t/\tau}$, with $P$ as the macroscopic polarization and $\tau$ as the macroscopic relaxation time. Just like the characteristic times of individual platelets, $\tau$ depends on many factors such as temperature, barium concentration, and chemical homogeneity.\\

As a consequence of the proposed switching mechanism, high permittivity can only occur if the platelets can first be excited into a chaotic state and subsequently relax into a quasistable state. This is most certainly possible in the temperature range $700~\mathrm{K} < T_\mathrm{sim} < 800~\mathrm{K}$ (i.e., $360~\mathrm{K} < T_\mathrm{exp} < 540~\mathrm{K}$), where both chaotic and quasistable stages occurred in the \textit{ab initio} MD simulations of the 001-ordered NBT. Indeed, NBT-3.6BT reaches the highest dielectric permittivity in this temperature range,\cite{Schneider2014} which clearly supports our hypothesis. Below $T_\mathrm{sim} = 700~\mathrm{K}$, no thermally induced chaos occurs, so that a polarized state does not decay. This corresponds to the ferroelectric behavior which is observed at $T_\mathrm{exp} < 450~\mathrm{K}$.\cite{Schneider2014} Above $T_\mathrm{sim} = 800~\mathrm{K}$, on the other hand, polarized states decay instantaneously when the applied field is removed, which corresponds to paraelectric behavior.\\

Further support can be derived from the time and frequency dependence. The observed duration of the chaotic stages does not exceed a few picoseconds. This corresponds to a relaxation frequency of about 100~GHz, which falls into the same range as the CC1 relaxation identified in pure NBT in Ref.~\onlinecite{Petzelt2014}. The slight shift in temperature and thus frequency is expected due to the addition of barium. \textcite{Petzelt2014} explain this CC1 relaxation as a bismuth ion hopping process, i.e., a dipole reorientation. Recalling the correlations between the bismuth and oxygen RMSD in Fig.\nolinebreak\ \ref{fig:RMSD+SCSF}, the bismuth hopping can easily be envisaged to occur concurrently with the previously discussed octahedral tilt relaxations.\\

\section{Conclusions}
The dielectric relaxation mechanism in NBT-3.6BT was investigated by QENS, and by \textit{ab initio} MD simulations on pure NBT with 001 and 111 cation order. The quasielastic part of the neutron scattering data could be reproduced by a linear combination of calculated spectra. The temperature scale had to be adjusted mainly due to the effects of barium addition. The 001-ordered NBT was associated with the tetragonal platelets in NBT-3.6BT due to the preferred c$^+$ tilting, whereas the 111-ordered NBT favors a$^-$a$^-$a$^-$ tilting and was associated with the rhombohedral matrix. The high 001 fraction necessary to reproduce the measured QENS spectra was therefore seen as an indication that the relaxational dynamics in NBT-3.6BT are concentrated in the tetragonal platelets.\\

Detailed analysis of the \textit{ab initio} MD trajectories confirmed a correlation of the bismuth and oxygen dynamics. As order parameters for in-phase and anti-phase tilting, the $\frac{1}{2}(310)$ and $\frac{1}{2}(311)$ structure factors, respectively, were used to characterize the different stages that can occur within one \textit{ab initio} MD simulation. Fundamental differences between quasistable and chaotic stages were demonstrated clearly. Furthermore, it was shown that relaxation processes on the timescale probed by our QENS experiments only occur during chaotic stages. In consequence, the hypothesis was developed that the macroscopic relaxor behavior results directly from the relaxation of the chaotic state in the tetragonal platelets. The distribution of relaxation times among the platelets might explain the macroscopic frequency dependence in the 100~GHz range. The paraelectric behavior of NBT-3.6BT at temperatures exceeding $T_\mathrm{exp} \approx 540~\mathrm{K}$ can be justified by the absence of quasistable stages, which are necessary to prevent the decay of an induced polarization.\\

Since the \textit{ab initio} MD simulations are limited to the picosecond timescale, and the QENS experiments were performed with an energy resolution around 1~meV, only the high frequency switching behavior could be investigated. The characteristic frequency dependence of the dielectric properties of NBT-3.6BT in the kHz to MHz range requires different experimental and theoretical treatment. Relaxation dynamics in the 100~kHz range could be measured with energy resolution around 1~neV or on the timescale of 10~$\mu$s. Simulations of these slow relaxations probably need to take domain switching processes into consideration as well, which occur on lengthscales far exceeding the box size of the \textit{ab initio} MD simulations presented above.\\

Furthermore, sound experimental proof of the correlation between chemical short range order and octahedral tilting would be desirable. The hypothetical switching mechanism presented above needs to be verified, too, for example using \textit{in situ} neutron scattering experiments or \textit{ab initio} MD simulations with applied electric field. The field direction, amplitude, and frequency are expected to have an influence on the results. Finally, it remains to be seen whether a similar switching mechanism can be identified in other relaxor systems, too.\\

\begin{acknowledgments}
The authors would like to acknowledge Daniel Rytz for the synthesis of the investigated single crystal. This work is based on experiments performed at the Swiss spallation neutron source SINQ, Paul Scherrer Institute, Villigen, Switzerland. The authors would also like to thank the Institut Laue-Langevin for the beamtime provided and the hospitality. This research project has been supported by the European Commission under the 7th Framework Programme through the `Research Infrastructures' action of the `Capacities' Programme, NMI3-II Grant number 283883. This work was funded by the Deutsche Forschungsgemeinschaft (DFG) within the SFB 595 `Electrical Fatigue in Functional Materials' and by the DFG Priority Programme 1599 (AL-578/16). The authors gratefully acknowledge the computing time granted by the John von Neumann Institute for Computing (NIC) and provided on the supercomputer JURECA\cite{Krause2016} at J\"ulich Supercomputing Centre (JSC). Moreover, computing time was granted on the Lichtenberg-High Performance Computer at TU Darmstadt.\end{acknowledgments}

\bibliography{QENS_PRB}

\end{document}